\documentclass[a4paper]{article}
\pdfoutput=1
\usepackage{jhepmod}

\usepackage[utf8]{inputenc}

\usepackage[table ]{ xcolor}
\usepackage{enumitem}
\usepackage{multirow}
\usepackage{xcolor}
\usepackage{amsmath}
\usepackage{amsthm}

\usepackage{tikz}
\usetikzlibrary{shapes,arrows}
\tikzstyle{startstop} = [rectangle,rounded corners, minimum width=3cm,minimum height=1cm,text centered, draw=black,fill=red!30]
\tikzstyle{io} = [trapezium, trapezium left angle = 70,trapezium right angle=110,minimum width=3cm,minimum height=1cm,text centered,draw=black,fill=red!30]
\tikzstyle{process} = [rectangle,minimum width=3cm,minimum height=1cm,text centered,text width =3cm,draw=black,fill=orange!30]
\tikzstyle{decision} = [diamond,minimum width=3cm,minimum height=1cm,shape aspect=3,inner sep = 0.4pt,text centered,draw=black,fill=green!30]
\tikzstyle{arrow} = [thick,->,>=stealth]
\tikzstyle{shadow}=[preaction={fill=black,opacity=.5,transform canvas={xshift=0.5mm,yshift=-0.5mm},shading=radial,shading angle=20},fill=red]

\tikzstyle{ellipse}=[draw, rectangle, minimum width=2.8em, rounded corners=6pt,line width=0.5pt]
\tikzstyle{pxsbx}=[trapezium, trapezium left angle=75, trapezium right angle=105, minimum width=3em, text centered, draw = black, fill=white,line width=0.5pt] 
\tikzstyle{lingxing}=[draw,diamond,shape aspect=3,inner sep = 0.4pt,thick,font=\itshape,line width=0.5pt]

\usepackage{amssymb}
\usepackage{graphicx,color}

\usepackage{amsmath,amsthm}

\usepackage{mathrsfs}
\usepackage{xcolor}

\usepackage{bm}

\def\beq{\begin{equation}}
\def\eeq{\end{equation}}
\newcommand{\bea}{\begin{eqnarray}}
\newcommand{\eea}{\end{eqnarray}}
\def\bi{\begin{itemize}}
\def\ei{\end{itemize}}
\def\ba{\begin{array}}
\def\ea{\end{array}}
\def\bfig{\begin{figure}}
\def\efig{\end{figure}}

\def\be{\begin{eqnarray}}
\def\ee{\end{eqnarray}}

\newcommand{\red}{\textcolor{red}}

\renewcommand{\o}{\omega}
\renewcommand{\O}{\Omega}

\newcommand{\lt}{\left}
\newcommand{\rt}{\right}

\title{Black hole tunneling in loop quantum gravity}

\author[1]{\ Hongwei Tan}

\author[2,3]{\ Rong-Zhen Guo}  

\author[4]{\ Jingyi Zhang}

\affiliation[1]{School of Science, Hunan Institute of Technology, Hengyang 421002, China}

\affiliation[2]{School of Fundamental Physics and Mathematical Sciences Hangzhou
Institute for Advanced Study, UCAS, Hangzhou 310024, China}
\affiliation[3]{School of Physical Sciences, University of Chinese Academy of Sciences, No. 19A Yuquan Road, Beijing 100049, China}
\affiliation[4]{Center for Astrophysics, Guangzhou University, Guangzhou 510006, China}

\emailAdd{Corresponding author:guorongzhen@ucas.ac.cn}
\emailAdd{htan2018@fau.edu}

\emailAdd{zhangjy@gzhu.edu.cn}

\abstract{
In this paper, we investigate the Hawking radiation of the quantum Oppenheimer-Snyde black hole  with the tunneling scheme by Parikh and Wilczek.
We calculate the emission rate of massless scalar particles.
Compared to the traditional results within the framework of General Relativity, our findings include quantum correction terms arising from loop quantum gravity effects.
Following the approach in \cite{zhang2008black,zhang2009black}, we establish the entropy of the black hole. This entropy  includes a logarithmic correction, which arises from quantum gravity effects.
Our result is consistent with the well-known result in the context of quantum gravity. }


\begin{document}

\maketitle

\section{Introduction}

As an important prediction  of General Relativity (GR), black hole (BH) have drawn considerable attentions and undergone widely study (see e.g., \cite{wald2010general, poisson2004relativist,frolov2011introduction,afshordi2024black,chandrasekhar1985mathematical}).
Both the gravitational wave detections \cite{LIGOScientific:2016aoc,LIGOScientific:2016sjg,LIGOScientific:2016dsl,LIGOScientific:2017bnn,abbott2016observation} and the observations of supermassive BHs by the Event Horizon Telescope \cite{event2019first, akiyama2019first, akiyama2019first-2,akiyama2019first-3,akiyama2019first-4,akiyama2019first-5}  provide the
strong evidences of the existence of BH. In the future, more gravitational wave and electromagnetic wave detection
will provide stronger tests of the BH paradigm \cite{Punturo:2010zz,Evans:2023euw,Evans:2021gyd,Colpi:2024xhw,TianQin:2015yph,Hu:2017mde}.
Despite the remarkable success in the classical paradigm of BH, there still some problems remain, e.g., the BH singularity and information paradox (see e.g., \cite{penrose1965gravitational,hawking1970singularities,brady1995black,hawking1974black,hawking2005black, bekenstein2004black,landgren2019information}), where quantum gravity potentially plays a significant role.

Loop quantum gravity (LQG), a background-independent and non-perturbative quantum gravity, is one candidate resolution of the problems mentioned previously \cite{thiemann2008modern, rovelli2004quantum,han2007fundamental,thiemann2003lectures,ashtekar2004background,giesel2012classical,rovelli2011zakopane,perez2013spin}.
By taking into account of the LQG effects, the BH singularity is replaced by a bounce, thus solving the singularity problem \cite{husain2022quantum,husain2022fate}.
Another advancement is in exploring the BH entropy and information paradox. LQG successfully reproduces BH entropy by counting the number of spacetime microstates \cite{rovelli1996black,perez2017black}. 
Recently, the concept of a black-to-white hole transition proposal is regraded as a promising candidate as the solution to the BH information paradox \cite{rovelli2014planck, haggard2015quantum,de2016improved,christodoulou2016planck,bianchi2018white,PhysRevD.103.106014,soltani2021end,rignon2022black,han2023geometry}. The addressing of these issues will be accompanied by effects beyond the standard BH paradigm. Therefore, LQG is expected to introduce phenomenological corrections to BHs within the framework of GR.

To obtain this modified model, studying the quantum corrections to the spherically symmetric self-gravitational collapse problem provides a promising starting point. The spherically symmetric gravitational collapse plays a crucial role in understanding  dynamical formation of BH, from both classical and quantum perspectives.
In classical theory, the first collapse model is proposed by Oppenheimer and Snyder \cite{oppenheimer1939continued}, known as the Oppenheimer-Snyder (OS) model.
This model assumes that the matter field in the interior region is a pressure-less dust field. Thus, the interior dynamics can be described by the standard Friedmann equations.
The exterior region is characterized by Schwarzschild solution according to Birkhoff's theorem. Due to its simplicity, this model can be solved exactly, offering valuable insights into the nature of self-gravitational collapse.

 In recent developments \cite{lewandowski2023quantum}, LQG effects have been incorporated into the classical OS model, i.e., quantum OS model in literature. It can be viewed as an effective model of LQG-modified BH. It is formed as the intermediate product of the self-gravitational collapse process. 
 Later, the collapsing phase transits to \textcolor{red}{an} expanding phase, resulting in  a black-to-white hole transition.
This BH has been extensively studied from various aspects, such as shadows and
quasinormal modes \cite{yang2023shadow,stashko2024quasinormal,zhang2023black,gong2024quasinormal,yang2024gravitational,liu2024gravitational,zi2024eccentric}.
 In this framework, the interior region is treated using the Ashtekar-Pawlowski-Singh (APS) model, where the classical Friedmann equation is modified by quantum corrections arising from loop quantum cosmology (LQC) effects. Appropriate junction conditions are then introduced to derive the spacetime metric in the exterior region. 
 Then the spacetime metric reads \footnote{Here we set $C=G=1$. In this article, the metric has the signature $(-,+,+,+)$.}
\begin{equation}
    \mathrm{d} s^2
    =-\left(1-\frac{2M}{r}+\frac{\alpha M^2}{r^4}\right)\mathrm{d}t^2
    +\left(1-\frac{2M}{r}+\frac{\alpha M^2}{r^4}\right)^{-1}\mathrm{d}r^2
    +r^2\mathrm{d}\O^2,
\end{equation}
where $M$ is the ADM energy of the spacetime,
and $\alpha=16 \sqrt{3} \pi \gamma^3 \ell_p^2$ with $\ell_p=\sqrt{\hbar}$ is the Planck length and $\gamma$ is the Barbero-Immirzi parameter.
Note that $\gamma$ is dimensionally less, $[\alpha]=[M]^2=[r]^2$.\footnote{$[A]$. denotes the dimension of the quantity [A].}
Compared to the classical Schwarzchild solution, a quantum correction arises from LQG effects.
We will discus this with more details later.

For simplicity, this paper focuses on the scenarios where the matter distribution in the interior spacetime region is assumed to be homogeneous and isotropic, in accordance with the cosmological principle. However, to better reflect realism, an anisotropic cosmological background for the interior spacetime region should also be considered. In particular, the Kantowski-Sachs spacetime is of critical importance in this context \cite{kantowski1966some,Joe:2015jis,modesto2006kantowski,casadio2024cosmology}.
This model describes the anisotropic collapse of the matter field, where expansion occurs in one direction while contraction takes place in the other directions.
We will investigate these topics in our future research, which may potentially provide new insights into the study of quantum gravity.

For a given BH, one can calculate its Hawking radiation and entropy. Historically, various methods have been developed for this purpose (see e.g., \cite{hawking1975particle,gibbons1977action,parikh2000hawking,parikh2004secret,rahman2012hawking,kim2016schwinger,hooft1985quantum,emparan2006black,almheiri2021entropy}).
One of them is the tunneling approach, a semiclassical method introduced by Parikh and Wilczek (PW) in \cite{parikh2000hawking, parikh2004secret}, and further developed in \cite{zhang2008black,Zhang:2005wn,Zhang:2005sf}.
In this approach, Hawking radiation is regarded as tunneling effects across the BH horizon,
with considerations for back-reaction effects of emitted particles and energy conservation of the system. This leads to a spectrum for emitted particles that is not purely thermal but includes a modification term. Additionally, the BH entropy can be derived from this spectrum.

In this work, we apply the PW approach to the quantum OS BH, focusing on massless scalar outgoing particles.
We compute the emission rate $\Gamma$ of the outgoing particles.
Comparing to the original result in classical GR
\cite{parikh2000hawking}, our results reveal additional quantum correction terms, 
which are interpreted as arising from quantum gravity effects.
Furthermore, by following the arguments in \cite{zhang2008black,zhang2009black}, we extract the BH entropy from the emission rate. Compared to the classical results, quantum correction terms also appear in the entropy formula for the quantum OS BH. By considering the quantum gravity effects,  quantum corrections contribute to the BH entropy, where a logarithmic term appears (see e.g., \cite{ghosh2005log,kaul2000logarithmic,domagala2004black,meissner2004black,chatterjee2004universal,medved2004comment,lin2024effective, shi2024higher,banerjee2008quantum,banerjee2008quantum2,banerjee2009quantum,majhi2009fermion,majhi2010hawking})
\begin{equation}\label{eq:lowboun_entropy}
    \tilde{S}= S+a\log \lt(\frac{{\mathscr{A}}}{l_p^2}\rt)+O(1).
\end{equation}
Here $a$ is a constant.
$\mathscr{A}$ is the area of the BH horizon.
$S$ is the classical BH entropy, given by $S=\frac{\mathscr{A}}{4l_p^2}$.
As the main result of our work, in the case of the quantum OS BH, we extract the BH entropy from the emission spectrum, which also includes a logarithmic correction
\begin{equation}
  \tilde{S}=S\red{+}\frac{\sqrt{2}\pi\alpha}{l_p^2}\log \lt(\frac{\mathscr{A}_{\text{Sch}}}{l_p^2}\rt)+O(\alpha^2).
\end{equation}
Here $\mathscr{A}_{\text {Sch}}$ is the area of the Schwarzschild BH horizon.
Our finding agrees with the traditional results \eqref{eq:lowboun_entropy}.
Besides, we obtain this result with a semiclassical approach, providing  evidence for the validity of  PW method.

This paper is organized as following: In Sec. \ref{sec:review_qOS}, we provide a brief review of quantum OS model. In Sec. \ref{sec:tunneling}, we compute  the emission rate of the  massless scalar particles created near the BH horizon and analyze the BH entropy.
In Sec. \ref{sec:con}, we provide conclusions and outlooks of this work.

\section{A brief review of quantum Oppenheimer-Snyder model}\label{sec:review_qOS}
Recently, the quantum OS model is investigated in \cite{lewandowski2023quantum,yang2023shadow,stashko2024quasinormal,zhang2023black,gong2024quasinormal,yang2024gravitational,liu2024gravitational}.
The classical  OS model is the first model which describes the collapse a pressure-less dust field and the formation of a BH in a spherically symmetric background \cite{oppenheimer1939continued, poisson2004relativist}. 
In this model, space is divided into two regions: the interior region $\mathcal{M}^-$, which is filled with the matter field, and the exterior $\mathcal{M}^+$, which is vacuum.
The geometry of the interior region is described by the usual FRW metric
\begin{equation}
    \mathrm{d} s_{\mathrm{APS}}^2=-\mathrm{d} \tau^2+a(\tau)^2\left(\mathrm{~d} \tilde{r}^2+\tilde{r}^2 \mathrm{~d} \Omega^2\right),
    \end{equation}
  where $\left(\tau,\,\tilde{r},\,\theta,\,\varphi\right)$ denotes a coordinate system.
  $a(\tau)$ is the scalar factor, which satisfies the classical Friedmann equation
  \begin{equation}
    H^2:=\left(\frac{\dot{a}}{a}\right)^2=\frac{8 \pi }{3} \rho,
    \end{equation}
    with $H$ is the Hubble constant.
The exterior region is vacuum, which is described by the Schwarzschild metric.
Ref. \cite{lewandowski2023quantum} considers the quantum version of the OS model.
The authors assume that the matter in the interior region satisfies a LQC-deformed Friedmann equation (see e.g., \cite{ashtekar2006quantum, yang2009alternative, assanioussi2018t})
\begin{equation}
    H^2:=\left(\frac{\dot{a}}{a}\right)^2=\frac{8 \pi }{3} \rho\left(1-\frac{\rho}{\rho_c}\right), \rho=\frac{M}{\frac{4}{3} \pi \tilde{r}_0^3 a^3}.
    \end{equation}
Here, $M$ is the ADM mass of the matter field.
The critical density is given  by $\rho_c=\sqrt{3} /\left(32 \pi^2 \gamma^3\hbar\right)$.
The corrected term $-\frac{8\pi\rho^2}{3\rho_c}$ arises from LQC effects.
In this work, we consider the mass of the matter to be on the scale of  solar mass.
The metric of the exterior region is given by applying appropriate junction conditions  to both the reduced metric and the extrinsic curvature of the junction surface $\Sigma$\footnote{The Latin letters $a,\,b,\,c$ are the abstract indices.}
\begin{equation}
    h_{ab}^+\left|_{\Sigma}\right.=h_{ab}^-\left|_{\Sigma}\right.,\quad
    K_{ab}^+\left|_{\Sigma}\right.=K_{ab}^-\left|_{\Sigma}\right..
\end{equation}
Here the "$+$" sign denotes the exterior region while the "$-$" sign denotes the interior region.
$\Sigma$ is a timelike hypersurface, which connects the exterior and the interior regions.
$h_{ab}$ is the reduced metric, while $K_{ab}$ is the extrinsic curvature of $\Sigma$.
In classical GR, the Israel junction conditions indicate that $K_{ab}$ exhibits a jump at $\Sigma$ only if the matter density is distributional \cite{israel1966singular}, meaning the matter possesses a surface density. However, in the quantum OS model, where the matter does not have a surface density, $K_{ab}$ is assumed to be continuous, as discussed in \cite{shi2024higher}.
The metric of the exterior region reads
\begin{equation}\label{eq:quan_OS}
    \mathrm{d} s^2
    =-\left(1-\frac{2M}{r}+\frac{\alpha M^2}{r^4}\right)\mathrm{d}t^2
    +\left(1-\frac{2M}{r}+\frac{\alpha M^2}{r^4}\right)^{-1}\mathrm{d}r^2
    +r^2\mathrm{d}\O^2.
\end{equation}
Compared to the Schwarzschild solution, there is a quantum-corrected term $\frac{\alpha M^2}{r^4}$ in eq. \eqref{eq:quan_OS}.
This term is interpreted as a contribution of LQG. The position of the BH horizon is given by 
\begin{equation}\label{eq:pos_of_hor}
    f(r)=1-\frac{2M}{r}+\frac{\alpha M^2}{r^4}=0.
\end{equation}
There are several roots in eq. \eqref{eq:pos_of_hor}.
However, as demonstrated in \cite{lewandowski2023quantum}, there are only two real roots when $M>M_{\text{min}}$ Here $M_{\text{min}}$ is the lower bound of the mass  required for BH formation, given by $M_{\min }:=\frac{4}{3 \sqrt{3} } \sqrt{\alpha}$. We focus only on the outer horizon, which reads
\begin{equation}
    \begin{aligned}
   r_h= \frac{M}{2}+\frac{1}{2} \sqrt{h}
    +\frac{1}{2} \sqrt{k},
    \end{aligned}
    \end{equation}
with 
\begin{equation}
   h=M^2+\frac{1}{3} M\left(\frac{2\cdot6^{2/3}\alpha}{\left(9 M \alpha+\sqrt{3} \sqrt{\left(27 M^2-16 \alpha\right) \alpha^2}\right)^{1 / 3}}+\left(54 M \alpha+6 \sqrt{3} \sqrt{\left(27 M^2-16 \alpha\right) \alpha^2}\right)^{1 / 3}\right),
\end{equation}
and
\begin{equation}
    \begin{aligned}
        k
        =&2 M^2-\frac{M\left(18 M \alpha+2 \sqrt{3} \sqrt{\left(27 M^2-16 \alpha\right) \alpha^2}\right)^{1 / 3}}{3^{2 / 3}}-\frac{2\cdot2^{2/3} M \alpha}{\left(27 M \alpha+3 \sqrt{3} \sqrt{\left(27 M^2-16 \alpha\right) \alpha^2}\right)^{1 / 3}}\\
        &+\frac{2 M^3}{\sqrt{M^2+\frac{1}{3} M\left(\frac{2 \cdot 6^{2 / 3} \alpha}{\left(9 M a+\sqrt{3} \sqrt{\left(27 M^2-16 \alpha\right) \alpha^2}\right)^{1 / 3}}+\left(54 M \alpha+6 \sqrt{3} \sqrt{\left(27 M^2-16 \alpha\right) \alpha^2}\right)^{1 / 3}\right)}}.
    \end{aligned}
\end{equation}
Note that $M^2\gg\alpha>\frac{16}{24}\alpha$, we find the reality of the square roots above is preserved. Also , we can employ Taylor expansion to $r_h$, which yields
\begin{equation}
    r_h=2M-\frac{\alpha}{8M}+O(\alpha^2)<2M.
\end{equation}
The term $-\frac{\alpha}{8M}$ is the quantum correction of LQG.
\section{Hawking radiation and black hole entropy}\label{sec:tunneling}
In this section, we investigate the Hawking radiation of quantum OS BH with the PW approach.
First, we rewrite the metric \eqref{eq:quan_OS} in Painlevé-Gullstrand coordinates \cite{painleve1921mecanique} and then compute the emission rate of the massless scalar particles near the horizon.
Finally, the BH entropy is discussed.
\subsection{Quantum Oppenheimer-Snyder black hole in Painlevé-Gullstrand coordinates}
To express the line element \eqref{eq:quan_OS} in Painlevé-Gullstrand coordinates, we introduce the following coordinate transformation
\begin{equation}
    \tilde{t}=t+F(r),
\end{equation}
where 
\begin{equation}
    F(r)=\int_0^{r}\sqrt{\frac{M r^4\left(2 r'^3-M \alpha\right)}{\left(-2 M r'^3+r'^4+M^2 \alpha\right)^2}}\mathrm{d}r'.
\end{equation}
The explicit expression is quiet complicated, but it is not important in our discussions. 
Then we find 
\begin{equation}\label{eq:quan_OS_pain}
    \mathrm{d} s^2
    =-\left(1-\frac{2M}{r}+\frac{\alpha M^2}{r^4}\right)\mathrm{d}\tilde t^2
    +2\sqrt{\frac{2M}{r}-\frac{\alpha M^2}{r^4}}\mathrm{d}\tilde t\mathrm{d}r
    +\mathrm{d}r^2
    +r^2\mathrm{d}\O^2.
\end{equation}
A brief review of Painlevé-Gullstrand coordinates and a detailed derivation of eq. \eqref{eq:quan_OS_pain} can be found in Appendix \ref{app:pain-coo}.
We observe that there is no coordinate singularity at the horizon in \eqref{eq:quan_OS_pain}, enabling us to investigate the tunneling process across the horizon.
Furthermore, the timeslice of \eqref{eq:quan_OS_pain} is Euclidean, allowing traditional Schr\"odinger equation is remain  valid.
\subsection{Parikh-Wilczek approach of quantum Oppenheimer-Snyder black hole}
In the PW approach \cite{parikh2000hawking}, the Hawking radiation is interpreted as a quantum tunneling process.
This approach takes into account the energy conservation condition and incorporates the backreaction of Hawking radiation.

We consider the emitted particles to be massless scalar particles.
The radial null geodesic reads
\begin{equation}
    \dot{r}=\pm 1-\sqrt{\frac{2M}{r}-\frac{\alpha M^2}{r^4}},
\end{equation}
with $\dot{r}$ denotes the derivative of $r$ with respective to $\tilde{t}$.
The "+" sign corresponds to an outgoing particle, while the "-" sign corresponds to an ingoing particle.
In the following discussion, we focus on the outgoing particle which tunnels across the outer BH horizon.
For simplicity, we assume the wave corresponding to the emitted particle is an s-wave.
In classical GR, Birkhoff's theorem state that any spherically symmetric solution of the vacuum field equations must be static and asymptotically flat.
Nevertheless, the case of quantum OS black hole is different, due to the contributions of  LQG effects.
Recent work \cite{cafaro2024status} addresses this issue by generalizing  Birkhoff's theorem within the context of Polymerized Einstein
Field Equations (PEFE), which incorporates quantum gravity corrections.
The authors demonstrate in this work that the exterior region of the quantum OS spacetime satisfies the generalized Birkhoff’s theorem, with three distinct branches corresponding to $k=-1,\,0$ and 1, representing open, flat and closed universes for the interior region, respectively. 
In this paper, we focus on the $k=0$ case, as shown in \eqref{eq:quan_OS}.
As demonstrated in [95], the exterior region of this type of quantum OS model is unique and characterized by a single parameter: the mass.
Therefore, no graviton is emitted in this process. 
Suppose the energy of this particle is $\omega$. According to the energy conservation, the energy of the BH changes as $M\to M-\o$.
Accordingly, the modified line element yields 
\begin{equation}\label{eq:quan_OS_pain_mod}
    \mathrm{d} s^2
    =-\left(1-\frac{2(M-\o)}{r}+\frac{\alpha (M-\o)^2}{r^4}\right)\mathrm{d}\tilde t^2
    +2\sqrt{\frac{2(M-\o)}{r}-\frac{\alpha (M-\o)^2}{r^4}}\mathrm{d}\tilde t\mathrm{d}r
    +\mathrm{d}r^2
    +r^2\mathrm{d}\O^2,
\end{equation}
and the modified radial null geodesic yields 
\begin{equation}\label{eq:null_geo_modi}
    \dot{r}=\pm 1-\sqrt{\frac{2(M-\o)}{r}-\frac{\alpha (M-\o)^2}{r^4}}.
\end{equation}
The tunneling process occurs very close to the horizon. The outgoing particle, as measured by the local observer near the horizon, experiences an ever-increasing blue shift. Consequently, the geometric optics approximation is suitable for describing the outgoing particles, and the WKB approximation is justified.

The outgoing particles are created inside the horizon. As they are emitted, they cross the horizon and travel to infinity. This process is classically forbidden but permitted through quantum tunneling. Next, we calculate the emission rate of these outgoing particles. 
With the WKB approximation, the emission rate is given by 
\begin{equation}
    \Gamma \sim e^{-\frac{2}{\hbar} \operatorname{Im} A},
\end{equation}
with $A$ is the action of the particle. The imaginary part of the action is given by 
\begin{equation}
    \operatorname{Im} A=\operatorname{Im} \int_{r_{\text {in }}}^{r_{\text {out }}} p_r \mathrm{d} r=\operatorname{Im} \int_{r_{\text {in }}}^{r_{\text {out }}} \int_0^{p_r} \mathrm{d}p_r^{\prime} \mathrm{d} r.
    \end{equation}
Here $p_r$ is the conjugate momentum of $r$, $r_{in}$ and $r_{out}$ are the locations of the particle before and after the tunneling across the event horizon. Hamiltonian equations imply
\begin{equation}
    \dot{r}=+\left.\frac{\mathrm{d} H}{\mathrm{d} p_r}\right|_r,
    \end{equation}
with the Hamiltonian is $H=M-\o$.
Then with eq. \eqref{eq:null_geo_modi} and focusing on the outgoing particles, we find
\begin{equation}\label{eq:ima_of_act}
    \operatorname{Im} A
    =\operatorname{Im} \int_{r_{\text {in }}}^{r_{\text {out }}} \int_0^{p_r} \mathrm{d} p_r^{\prime} \mathrm{d} r
    =-\operatorname{Im} \int_{r_{\text {in }}}^{r_{\text {out }}} \int_0^{\o}\frac{\mathrm{d}r}{1-\sqrt{\frac{2(M-\o')}{r}-\frac{\alpha (M-\o')^2}{r^4}}} \mathrm{d}\o'.
    \end{equation}
As assumed previously, $M^2\gg\alpha$, we can apply Taylor expansion to the integrand of eq. \eqref{eq:ima_of_act}
\begin{equation}\label{eq:tylor_expan}
    \frac{1}{1-\sqrt{\frac{2(M-\o')}{r}-\frac{\alpha (M-\o')^2}{r^4}}} 
    =\frac{1}{1-\sqrt{\frac{2(M-\o')}{r}}}
    -\frac{(M-\o')\sqrt{\frac{(M-\o')}{r}}}{2\sqrt{2}r^3\lt(1-\sqrt{\frac{2(M-\o')}{r}}\rt)^2}\alpha
    +O\lt(\alpha^2\rt).
\end{equation}
Note that $[\alpha]=[M]^2$. Hence, the left-hand side of \eqref{eq:tylor_expan} is dimensionless, which implies that any term in the right hand side of \eqref{eq:tylor_expan} is also dimensionless. Especially, $\left[\frac{(M-\o')\sqrt{\frac{(M-\o')}{r}}}{2\sqrt{2}r^3\lt(1-\sqrt{\frac{2(M-\o')}{r}}\rt)^2}\alpha\right]=1$.
The integrand is singular at the location of the classical horizon. To evaluate eq. \eqref{eq:ima_of_act}, we apply Feynman’s scheme. To ensure that the positive energy solutions decay over time, we deform the contour as $\omega \to \omega - i\epsilon$ with $\epsilon \to 0$. By deforming the contour, the singularity in eq. \eqref{eq:tylor_expan} is avoided, ensuring that the right-hand side of eq. \eqref{eq:tylor_expan} converges uniformly. Consequently, the integration and summation can be interchanged.
Then we find
\begin{equation}
    \begin{aligned}
        &\operatorname{Im} A\\
    =&-\lim_{\epsilon\to0}\operatorname{Im} \int_{r_{\text {in }}}^{r_{\text {out }}} \int_0^{\o}
    \left(
        \frac{1}{1-\sqrt{\frac{2(M-\o'+i\epsilon)}{r}}}
    -\frac{(M-\o'+i\epsilon)\sqrt{\frac{(M-\o'+i\epsilon)}{r}}}{2\sqrt{2}r^3\lt(1-\sqrt{\frac{2(M-\o'+i\epsilon)}{r}}\rt)^2}\alpha
    \right)\mathrm{d}r\mathrm{d}\o'
    +O(\alpha^2).
    \end{aligned}
\end{equation}
Introducing the coordinate transformation $u=\sqrt{r}$, we find
\begin{equation}\label{eq:int_S_final}
    \begin{aligned}
       & \operatorname{Im} A\\
    =&-\lim_{\epsilon\to0}\operatorname{Im} \int_{u_{\text {in }}}^{u_{\text {out }}} \int_0^{\o}
    \left(
        \frac{2u^2}{u-\sqrt{2(M-\o'+i\epsilon)}}
    -\frac{(M-\o'+i\epsilon)^{3/2}}{\sqrt{2}u^4\lt(u-\sqrt{2(M-\o'+i\epsilon)}\rt)^2}\alpha
    \right)\mathrm{d}u\mathrm{d}\o'\\
    &+O(\alpha^2).
    \end{aligned}
\end{equation}
We focus on the particles created near the outer horizon.
There is a pole at $u=\sqrt{2(M-\o'+i\epsilon)}$ in eq. \eqref{eq:int_S_final}.
Therefore, we find 
\begin{equation}
    \begin{aligned}
       \operatorname{Im} A
    =\lim_{\epsilon\to0}\int_0^{\o}
    \left(
       4\pi (M-\o')
    +\operatorname{Im}\frac{\sqrt{2}\pi i}{M-\o'+i\epsilon}\alpha
    \right)\mathrm{d}\o'
    &+O(\alpha^2).
    \end{aligned}
\end{equation}
For a solar mass black hole, we have $M\gg\o$.
Therefore,
\begin{equation}
    \begin{aligned}
        \operatorname{Im} A
    =&4\pi M\o\left(1-\frac{\o}{2M}\right)
    +\sqrt{2}\pi\alpha\left(\log \lt(\frac{M}{l_p}\rt)-\log\lt(\frac{M-\o}{l_p}\rt)\right)
    +O\left(\alpha^2\right)\\
    =&4\pi M\o\left(1-\frac{\o}{2M}\right)
    +\frac{\sqrt{2}}{2}\pi\alpha\left(\log \lt(\frac{\mathscr{A}_{\text{Sch}}\lt(M\rt)}{l_p^2}\rt)-\log\lt(\frac{\mathscr{A}_{\text{Sch}}\lt(M-\o\rt)}{l_p^2}\rt)\right)
    +O\left(\alpha^2\right)
    \end{aligned}.
\end{equation}
Here 
$\textcolor{red}{\mathscr{A}}_{\text {Sch} }$ is the area of the Schwarzschild BH horizon with mass $M$.
Compared with the original work \cite{parikh2000hawking}, an additional term $\frac{\sqrt{2}}{2}\pi\alpha\left(\log \lt(\frac{\mathscr{A}_{\text{Sch}}\lt(M\rt)}{l_p^2}\rt)-\log\lt(\frac{\mathscr{A}_{\text{Sch}}\lt(M-\o\rt)}{l_p^2}\rt)\right)$ appears.
This term is interpreted as the quantum correction arising from LQG.
Then the emisson rate of the particles yields
\begin{equation}\label{eq:emiss_rate_OS}
    \begin{aligned}
        &\Gamma\sim\exp(-\frac{2}{\hbar}\operatorname{Im} A)\\
        =&\exp\left(-\frac{8\pi M\o}{l_p^2}\left(1-\frac{\o}{2M}\right)
+\frac{\sqrt{2}\pi\alpha}{l_p^2}\left(\log \lt(\frac{\mathscr{A}_{\text{Sch}}\lt(M-\o\rt)}{l_p^2}\rt)-\log\lt(\frac{\mathscr{A}_{\text{Sch}}\lt(M\rt)}{l_p^2}\rt)\right)+O\left(\alpha^2\right)\right)
    \end{aligned}
\end{equation}
\subsection{The entropy of the quantum Oppenheimer-Snyde black hole}
In classical GR, the emission rated can be expressed as 
\begin{equation}\label{eq:emiss_in_entro}
    \Gamma\sim\exp{\Delta S},
\end{equation}   
with $\Delta S$ is the difference  of the black hole entropy between before and after the emission of the particle.
It is stated that this result implies the unitary of black hole evaporation in GR \cite{zhang2009black,zhang2008black}.
In  quantum mechanics, the transition apmplitude between the initial and the final states is 
\begin{equation}
    \Gamma(i\to f)
    =\left|\mathcal{M}\right|^2 \cdot(\text { phase space factor) },
\end{equation}
with $\left|\mathcal{M}\right|$ is the probabilty magnitude of this poccess.
The phase space factor is given by 
\begin{equation}
    \text { phase space factor }=\frac{N_f}{N_i}=\frac{e^{S_f}}{e^{S_i}}=e^{\Delta S}.
    \end{equation}
Let's take the Schwarzschild BH as an example.
For a Schwarzschild BH, its initial entropy is given by 
\begin{equation}
S_i=\frac{4\pi M^2}{l_p^2}.
\end{equation}  
After the black hole emits particle with energy $\o$, the final entropy is 
\begin{equation}
    S_f=\frac{4\pi (M-\o)^2}{l_p^2}.
\end{equation} 
Therefore, 
\begin{equation}
    \Delta S= S_f-S_i=-\frac{8\pi M\o}{l_p^2}\left(1-\frac{\o}{2M}\right)
    =-\frac{2}{\hbar} \operatorname{Im} A_{\text {Sch }},
\end{equation}
with $A_{\text {Sch }}$ is the action of the particle created near the Schwarzschild BH horizon.
Hence, the emission rate of this particle is 
\begin{equation}
    \Gamma_{\text {Sch }}\sim\exp{\Delta S}.
\end{equation} 
We now turn back to the cases of quantum OS BH.
Quantum correction terms appear in eq. \eqref{eq:emiss_rate_OS}.
Hence, to express the emission rate $\Gamma$ in the formula of eq.\eqref{eq:emiss_in_entro}, we introduce the quantum-corrected entropy as 
\begin{equation}\label{eq:modi_entro}
    \tilde{S}=S\red{+}\frac{\sqrt{2}\pi\alpha}{l_p^2}\log \lt(\frac{\textcolor{red}{\mathscr{A}}_{\text{Sch}}}{l_p^2}\rt)+O(\alpha^2),
\end{equation}
Here $S$ is the entropy of the Schwarzschild BH, given by $S=\frac{4\pi M^2}{l_p^2}$.
Note that we are considering particles  tunneling across the outer BH horizon.
Therefore, $\tilde{S}$ is the entropy of the outer BH horizon, Our approach aligns with the scheme introduced in \cite{parikh2000hawking,Zhang:2005wn,Zhang:2005sf}.
To compute the total BH entropy, a multi-horizon scenario should be considered  \cite{singha2022thermodynamics,saghafi2023hawking}.
This approach is more intricate than it may initially appear, and we will explore this topic further in our future research.
Finally, the emission rate reads
\begin{equation}
    \Gamma\sim\exp\left[\tilde S_f-\tilde S_i\right]
    =\exp{\Delta\tilde S}.
\end{equation}  
The pre-factor of the  logarithmic correction of the BH entropy relates to the micro-states of BH horizon.
In our findings, the sign of this pre-factor is positive. Although this differs from the results obtained within the framework of ordinary LQG \cite{ghosh2005log,kaul2000logarithmic,domagala2004black}, it is consistent with the entropy of the effective loop quantum BH computed by other approaches \cite{lin2024effective, shi2024higher}.
Ref. \cite{medved2004comment} provides profound insights into this discrepancy.
On the one hand, it states that the quantum correction to the BH entropy, arising from quantum gravity effects, reduces the entropy. On the other hand, it states that the logarithmic correction, which originates from thermal fluctuations, increases the entropy due to the enhanced uncertainty introduced by these fluctuations.
Based on these insights, our finding \eqref{eq:modi_entro} might suggest the logarithmic correction for quantum OS BH entropy incorporates the contributions from both quantum gravity effect and (effective) thermal fluctuations.
For precise, let us denote $a_q$ as the pre-factor for the logarithmic correction contributed by quantum gravity effect and $a_f$ as the pre-factor for the logarithmic correction contributed by (effective) thermal fluctuation, then $a_q+a_f=\frac{\sqrt{2}\pi\alpha}{l_p^2}$.
We will continue this interesting topic in our future researches.
Moreover, it has been reported that the BH entropy is related to the dimension of the spacetime \cite{nozari2007existence}.
Our finding is consistent with the situation of 4-dimensional spacetime discussed in \cite{nozari2007existence}.
Unlike the arguments based on quantum gravity, we get this correction with semiclassical method.
Therefore, our result supports the validity of the semiclassical method in the context of quantum gravity.
\subsection{Comparison with other scenarios for computing black bole entropy}
In addition to the previously introduced PW approach, several other methods for computing black hole entropy incorporate correction terms that modify the standard entropy formula. In this subsection, we compare these methods with our results.
\begin{itemize}
    \item \textbf{Noncommutative spacetime geometry approach}\\
    This approach is based on the assumption that the coordinates for the spacetime manifold do not commute to each other but instead satisfy the following relationship\footnote{The Greek letters $\mu,\,\nu$ are the abstract indices, which take the value $\mu,\,\nu=0,\,1,\,2,\,3$.}
    \begin{equation}
    \left[x^\mu,x^\nu\right]=i\theta^{\mu\nu},
    \end{equation}
    where $\theta^{\mu\nu}$  is an antisymmetric matrix that encodes the noncommutativity of spacetime \cite{snyder1947quantized,seiberg1999string,douglas2001noncommutative}. 
    The spherical and static solution within this framework reads \cite{aschieri2005gravity,nozari2008hawking}
    \begin{equation}
d s^2=-\left(1-\frac{2 M_\theta}{r}\right) \mathrm{d}t^2+\left(1-\frac{2 M_\theta}{r}\right)^{-1} \mathrm{d} r^2+r^2 \mathrm{d} \Omega^2.
\end{equation}
Here, $M_\theta$ is the mass of the spacetime satisfying the Gaussian distribution of minimal width $\sqrt{\theta}$
\begin{equation}
M_\theta=\int_0^r \rho_\theta(r) 4 \pi r^2\mathrm{d} r
=\frac{2 M}{\sqrt{\pi}} \int_0^{\frac{r^2}{4 \theta}} x^{\frac{1}{2}} e^{-x} \mathrm{d} x,
\end{equation}
with $M$ is a constant and the distribution function $\rho(r)$ reads
\begin{equation}\label{eq:modi_sch}
\rho_\theta(r)=\frac{M}{(4 \pi \theta)^{\frac{3}{2}}} \exp \left(-\frac{r^2}{4 \theta}\right).
\end{equation}
One can show that 
\begin{equation}
\lim _{\theta \rightarrow 0} M_\theta=M.
\end{equation}
In this case,  eq. \eqref{eq:modi_sch} reduces to he ordinary Schwarzschild solution.
Within this approach, the BH entropy entropy reads
\begin{equation}
    S_{\text{NSG}}=4 \pi \frac{M^2}{l_p^2} \mathcal{E}\left(\frac{M}{l_p\sqrt{\theta}}\right)-6 \pi \theta \mathcal{E}\left(\frac{M}{l_p\sqrt{\theta}}\right)+12 \sqrt{\pi \theta} \frac{M}{l_p} \exp \left(-\frac{M^2}{l_p^2\theta}\right),
\end{equation}
with $\epsilon\left(x\right)$ is the Gauss error function
\begin{equation}
\mathcal{E}(x) \equiv \frac{2}{\sqrt{\pi}} \int_0^x e^{-\lambda^2} \mathrm{d} \lambda.
\end{equation}
Especially, in the large mass limit $\frac{M}{l_p\sqrt{\theta}} \gg 1$, one finds 
\begin{equation}\label{eq:entro_noncom}
    S_{\text{NSG}}= \frac{4 \pi M^2}{l_p^2}+12 \sqrt{\pi \theta} \frac{M}{l_p}\exp \left(-\frac{M^2}{l_p^2\theta}\right).
\end{equation}
The  correction in eq. \eqref{eq:entro_noncom} is exponential, differing from the logarithmic correction in our result \eqref{eq:modi_entro}.
In fact, Ref. \cite{nozari2008hawking} demonstrates the BHs cannot evaporate completely in the noncommutative spacetime geometry framework. 
Instead, there is a lower bound for the mass, $M_0$, during the later stages of BH evaporation, and the information can be stored in the remnant, offering a possible resolution to the black hole information paradox. 
Our findings may lead to a similar conclusion, due to the pre-factor of the logarithmic correction in eq. \eqref{eq:modi_entro} is positive.
However, a subtlety arises because our analysis assumes  $M^2\gg\alpha$, which may not hold during the later stages of evaporation.
This suggests that these two frameworks might describe different phases of black hole evaporation. Investigating the relationship between these two approaches would be an interesting avenue for future research.
Moreover, As discussed in \cite{lewandowski2023quantum},  a bounce occurs at the end of collapse.
Exploring how to incorporate both the bounce and black hole evaporation into a unified framework remains a challenging but intriguing task.
\item \textbf{Modified dispersion relations approach}\\
Inspired by quantum gravity , the modified dispersion relations (MDR) approach arises from modifying energy-momentum dispersion relation \cite{amelino2006black, nozari2006comparison}
\begin{equation}\label{eq:modi_disp_relat}
\vec{p}^2= E^2-\mu^2+\alpha_1 l_p E^3+\alpha_2 l_p^2 E^4
+\alpha_3l_p^3E^5
+\alpha_4l_p^4E^6
+O\left(l_p^5 E^7\right),
\end{equation}
with $ E \ll 1 / l_p$.
The value of the coefficients $\alpha_i$ depend on the specific quantum gravity theory.
Applying the modified dispersion relation, the BH entropy is given by
\begin{equation}
\begin{aligned}
S_{\text{MDR}} \simeq & \frac{\mathscr{A}}{4 l_p^2}+\frac{\alpha_1 \pi^{\frac{1}{2}}}{l_p} \mathscr{A}^{\frac{1}{2}}+\pi\left(\frac{3}{2} \alpha_2-\frac{3}{8} \alpha_1^2\right) \log \frac{\mathscr{A}}{l_p^2} \\
& -4 \pi^{\frac{3}{2}} l_p\left(-\alpha_1 \alpha_2+\frac{1}{4} \alpha_1^3+2 \alpha_3\right) \mathscr{A}^{\frac{-1}{2}} \\
& -4 \pi^2 l_p^2\left(-\frac{5}{4} \alpha_1 \alpha_3-\frac{5}{8} \alpha_2^2+\frac{15}{16} \alpha_1^2 \alpha_2-\frac{25}{128} \alpha_1^4\right)\mathscr{A}^{-1} \\
& -\frac{16}{3} \pi^{\frac{5}{2}} l_p^3\left(\frac{9}{8} \alpha_1^2 \alpha_3-\frac{45}{48} \alpha_1^3 \alpha_2+\frac{9}{8} \alpha_1 \alpha_2^2\right. \\
& \left.+\frac{21}{128} \alpha_1^5-\frac{3}{2} \alpha_2 \alpha_3\right) \mathscr{A}^{-\frac{3}{2}} .
\end{aligned}
\end{equation}
As observed in \cite{nozari2006comparison},if the odd powers of energy in \eqref{eq:modi_disp_relat} are set to  0 ($\alpha_1=\alpha_3=0$), then the BH entropy reduces to 
\begin{equation}
S_{\text{MDR}} \simeq \frac{\mathscr{A}}{4 l_p^2}+\frac{3}{2} \pi \alpha_2 \log \frac{\mathscr{A}}{l_p^2}+\frac{5}{2} \pi^2 \alpha_2^2 \frac{l_p^2}{\mathscr{A}}.
\end{equation}
This result is consistent with our finding \eqref{eq:modi_entro}.
In fact, BH entropy formula imposes  constraints on the form of MDR.
\item  \textbf{Generalized uncertainty principle approach}\\
The general uncertainty principle (GUP) approach is based on the modification of the ordinary eisenbergnuncertainty principle (HUP) \cite{amelino2006black,nozari2012natural}
\begin{equation}
\delta x \geq \frac{1}{\delta p}+\beta l_p^2 \delta p+O\left(l_p^3 \delta p^2\right),
\end{equation}
with $\beta$ is a small parameter depends on the specific quantum gravity theory.
Based on this assumption, the BH entropy is computed as 
\begin{equation}
S_{\text{GUP}} \simeq \frac{\mathscr{A}}{4 l_p^2}-\beta \pi \ln \frac{\mathscr{A}}{l_p^2}.
\end{equation}
This result is consistent with our finding \eqref{eq:modi_entro} up to a difference in the sign.
Furthermore, by considering higher order effects, Ref. \cite{nozari2012natural} derives the BH entropy with additional  corrections and predicts that the existence of a remnant as the BH mass $M$ approaches to the order of  Planck mass $M_p$ during evaporation. 
 This prediction aligns with the results of the noncommutative spacetime geometry approach. Incorporating the GUP framework into the quantum OS BH model would be an intriguing direction for future research.
\item \textbf{Polymeric quantization approach}\\
The polymeric quantization approach is a method for quantizing gravity, inspired by techniques from LQG and quantum mechanics in the polymer representation\cite{ashtekar2003quantum,corichi2007polymer}.
It is based on the premise that, in a quantum theory of gravity, spacetime is discrete and there exists a minimum measurable length scale in the order of the Planck length. Within this framework, the classical phase space is modified to incorporate discrete structures, which is different from the standard continuous treatment.
Ref. \cite{gorji2014polymeric} computes the BH entropy in this framework.
We begin by introducing the polymeric area of the black hole in terms of the horizon area, expressed as:
\begin{equation}
\mathscr A^{\text {(poly) }}=\mathscr A\left(1-\frac{\left(1+\mu^2 M_P / 8\right)}{8 \pi}\left(\frac{M_P}{M }\right)^2\right)^2.
\end{equation}
Here, $\mu=\mu_0/\hbar$, with $\mu_0$ is the discreteness parameter.
The continuum limit is given by $\mu\to0$.
In the high-temperature limit, the black hole entropy is given by
\begin{equation}
S_{\text {poly }}=\frac{\mathscr A^{(\text {poly })}}{4 l_P^2}-\frac{1}{2} \ln \left[\frac{\mathscr A^{(\text {poly })}}{4 l_P^2}\right]+M \left(1-\frac{\mathscr A^{(\text {poly })}}{\mathscr A}\right)+\mathcal{O}\left[\mathscr A^{(\text {poly })^{-1}}\right].
\end{equation}
In the limit $\frac{M_p}{M}\to0$, the polymeric area $\mathscr A^{(\text {poly })}$ reduces to the original horizon area $\mathscr A$. 
The entropy reduces to the well-known form
\begin{equation}
S_{\text {poly }}=\frac{\mathscr A}{4 l_P^2}-\frac{1}{2} \ln \left[\frac{\mathscr A}{4 l_P^2}\right]+\mathcal{O}\left[\mathscr A^{-1}\right].
\end{equation}
This result is also consistent with our finding \eqref{eq:modi_entro}, up to a minus sign.
\end{itemize}

\section{Conclusion and outlook}\label{sec:con}
In this work, we apply the the PW approach to the quantum OS BH
and evaluate the emission rate of the outgoing massless scalar particles.
Compared to the original results in \cite{parikh2000hawking,parikh2004secret},
our findings include quantum correction terms, arising from LQG effects.
Following the scheme in \cite{zhang2009black,zhang2008black}, we establish the entropy of the OS BH.
This entropy formula includes a logarithmic correction, which is consistent with well-known result in the context of quantum gravity \cite{ghosh2005log,kaul2000logarithmic,domagala2004black}.

So far, we have only considered the massless scalar particles as the emitted particles.
It would be interesting to extend the study to the tunneling process of massive particles. 
This topic has been explored within the framework of classical GR in
\cite{zhang2008black,zhang2009black,Zhang:2005sf},
where BH thermodynamics plays a crucial role.
Investigating the tunneling process of massive particles in the OS model potentially provides us deeper insight into the BH thermodynamics in the context of LQG.

Recently, the island scheme is  proposed as a resolution to the BH information paradox \cite{penington2020entanglement,almheiri2019entropy,almheiri2020page,almheiri2019islands,hashimoto2020islands}.
This approach utilizes the concepts of the minimal
quantum extremal surface \cite{ryu2006aspects,hubeny2007covariant,engelhardt2015quantum} to evaluate the BH entropy,
successfully recovering the Page curve \cite{page1993information,page2013time} and resolving the BH information paradox.
In our future work, we plan to extend the island scheme to the OS BH and compare the results with those presented in this paper.
\section{Acknowledgments}
This work was supported by the National Natural Science Foundation of China under Grant No. 11873025.
\appendix
\section{ A brief review of Painlevé-Gullstrand coordinates}\label{app:pain-coo}
In this appendix, we briefly review the scheme of obtaining the Painlevé-Gullstrand coordinate transformation of a general static spacetime.
Firstly, we introduce the line element as 
\begin{equation}
    \mathrm{d}s^2=-\left(1-g(r)\right)\mathrm{d}t^2
    +\frac{1}{1-g(r)}\mathrm{d}r^2
    +\mathrm{d}\O^2,
\end{equation}
with $g(r)$ is a function of $r$.
We then introduce the coordinate transformation
\begin{equation}
    t=\tilde{t}+F(r).
\end{equation}
So we have 
\begin{equation}
    \mathrm{d}t=\mathrm{d}\tilde{t}+F'(r)\mathrm{d}r.
\end{equation}
In  Painlevé\textcolor{red}{-Gullstrand} coordinates, the time slice is required to be Euclidean,
which necessitates the coefficient of $\mathrm{d}r^2$ is 1.
consequently, we obtain the following equations for $F(r)$
\begin{equation}
    \frac{1}{1-g(r)}-[1-g(r)]\left[F^{\prime}(r)\right]^2=1.
    \end{equation}
Then the new line element reads 
\begin{equation}
    \mathrm{d} s^2=-\lceil 1-g(r)] \mathrm{d} t^2 \pm 2 \sqrt{g(r)} \mathrm{d} t \mathrm{~d} r+\mathrm{d} r^2+r^2 \mathrm{~d} \Omega^2.
    \end{equation}
In our case,  
\begin{equation}
    g(r)=\frac{2M}{r}-\frac{\alpha M^2}{r^4}.
\end{equation}
Then $F(r)$ is given by 
\begin{equation}
    F(r)=\int_0^{r}\sqrt{\frac{M r^4\left(2 r'^3-M a\right)}{\left(-2 M r'^3+r'^4+M^2 \alpha\right)^2}}\mathrm{d}r'.
\end{equation}
\bibliographystyle{jhep}
\bibliography{paper}

\providecommand{\href}[2]{#2}\begingroup\raggedright\begin{thebibliography}{100}

\bibitem{zhang2008black}
J.~Zhang, {\it Black hole quantum tunnelling and black hole entropy correction},  {\em Physics Letters B} {\bf 668} (2008), no.~5 353--356.

\bibitem{zhang2009black}
J.~Zhang, {\it Black hole entropy, log correction and inverse area correction},  {\em Physics Letters B} {\bf 675} (2009), no.~1 14--17.

\bibitem{wald2010general}
R.~M. Wald, {\em General relativity}.
\newblock University of Chicago press, 2010.

\bibitem{poisson2004relativist}
E.~Poisson, {\em A relativist's toolkit: the mathematics of black-hole mechanics}.
\newblock Cambridge university press, 2004.

\bibitem{frolov2011introduction}
V.~P. Frolov and A.~Zelnikov, {\em Introduction to black hole physics}.
\newblock OUP Oxford, 2011.

\bibitem{afshordi2024black}
N.~Afshordi, A.~Ashtekar, E.~Barausse, E.~Berti, R.~Brito, L.~Buoninfante, R.~Carballo-Rubio, V.~Cardoso, G.~Carullo, M.~Dafermos, et~al., {\it Black holes inside and out 2024: visions for the future of black hole physics},  {\em arXiv preprint arXiv:2410.14414} (2024).

\bibitem{chandrasekhar1985mathematical}
S.~Chandrasekhar and K.~S. Thorne, {\it The mathematical theory of black holes},  1985.

\bibitem{LIGOScientific:2016aoc}
{\bf LIGO Scientific, Virgo} Collaboration, B.~P. Abbott et~al., {\it {Observation of Gravitational Waves from a Binary Black Hole Merger}},  {\em Phys. Rev. Lett.} {\bf 116} (2016), no.~6 061102, [\href{http://arxiv.org/abs/1602.03837}{{\tt arXiv:1602.03837}}].

\bibitem{LIGOScientific:2016sjg}
{\bf LIGO Scientific, Virgo} Collaboration, B.~P. Abbott et~al., {\it {GW151226: Observation of Gravitational Waves from a 22-Solar-Mass Binary Black Hole Coalescence}},  {\em Phys. Rev. Lett.} {\bf 116} (2016), no.~24 241103, [\href{http://arxiv.org/abs/1606.04855}{{\tt arXiv:1606.04855}}].

\bibitem{LIGOScientific:2016dsl}
{\bf LIGO Scientific, Virgo} Collaboration, B.~P. Abbott et~al., {\it {Binary Black Hole Mergers in the first Advanced LIGO Observing Run}},  {\em Phys. Rev. X} {\bf 6} (2016), no.~4 041015, [\href{http://arxiv.org/abs/1606.04856}{{\tt arXiv:1606.04856}}]. [Erratum: Phys.Rev.X 8, 039903 (2018)].

\bibitem{LIGOScientific:2017bnn}
{\bf LIGO Scientific, VIRGO} Collaboration, B.~P. Abbott et~al., {\it {GW170104: Observation of a 50-Solar-Mass Binary Black Hole Coalescence at Redshift 0.2}},  {\em Phys. Rev. Lett.} {\bf 118} (2017), no.~22 221101, [\href{http://arxiv.org/abs/1706.01812}{{\tt arXiv:1706.01812}}]. [Erratum: Phys.Rev.Lett. 121, 129901 (2018)].

\bibitem{abbott2016observation}
B.~P. Abbott, R.~Abbott, T.~Abbott, M.~Abernathy, F.~Acernese, K.~Ackley, C.~Adams, T.~Adams, P.~Addesso, R.~X. Adhikari, et~al., {\it Observation of gravitational waves from a binary black hole merger},  {\em Physical review letters} {\bf 116} (2016), no.~6 061102.

\bibitem{event2019first}
E.~H.~T. Collaboration et~al., {\it First m87 event horizon telescope results. i. the shadow of the supermassive black hole},  {\em arXiv preprint arXiv:1906.11238} (2019).

\bibitem{akiyama2019first}
K.~Akiyama, A.~Alberdi, W.~Alef, K.~Asada, R.~Azulay, A.-K. Baczko, D.~Ball, M.~Balokovi{\'c}, J.~Barrett, D.~Bintley, et~al., {\it First m87 event horizon telescope results. ii. array and instrumentation},  {\em The Astrophysical Journal Letters} {\bf 875} (2019), no.~1 L2.

\bibitem{akiyama2019first-2}
K.~Akiyama, A.~Alberdi, W.~Alef, K.~Asada, R.~Azulay, A.-K. Baczko, D.~Ball, M.~Balokovi{\'c}, J.~Barrett, D.~Bintley, et~al., {\it First m87 event horizon telescope results. iii. data processing and calibration},  {\em The Astrophysical Journal Letters} {\bf 875} (2019), no.~1 L3.

\bibitem{akiyama2019first-3}
K.~Akiyama, A.~Alberdi, W.~Alef, K.~Asada, R.~Azulay, A.-K. Baczko, D.~Ball, M.~Balokovi{\'c}, J.~Barrett, D.~Bintley, et~al., {\it First m87 event horizon telescope results. iv. imaging the central supermassive black hole},  {\em The Astrophysical Journal Letters} {\bf 875} (2019), no.~1 L4.

\bibitem{akiyama2019first-4}
K.~Akiyama, A.~Alberdi, W.~Alef, K.~Asada, R.~Azulay, A.-K. Baczko, D.~Ball, M.~Balokovi{\'c}, J.~Barrett, D.~Bintley, et~al., {\it First m87 event horizon telescope results. v. physical origin of the asymmetric ring},  {\em The Astrophysical Journal Letters} {\bf 875} (2019), no.~1 L5.

\bibitem{akiyama2019first-5}
K.~Akiyama, A.~Alberdi, W.~Alef, K.~Asada, R.~Azulay, A.-K. Baczko, D.~Ball, M.~Balokovi{\'c}, J.~Barrett, D.~Bintley, et~al., {\it First m87 event horizon telescope results. vi. the shadow and mass of the central black hole},  {\em The Astrophysical Journal Letters} {\bf 875} (2019), no.~1 L6.

\bibitem{Punturo:2010zz}
M.~Punturo et~al., {\it {The Einstein Telescope: A third-generation gravitational wave observatory}},  {\em Class. Quant. Grav.} {\bf 27} (2010) 194002.

\bibitem{Evans:2023euw}
M.~Evans et~al., {\it {Cosmic Explorer: A Submission to the NSF MPSAC ngGW Subcommittee}},  \href{http://arxiv.org/abs/2306.13745}{{\tt arXiv:2306.13745}}.

\bibitem{Evans:2021gyd}
M.~Evans et~al., {\it {A Horizon Study for Cosmic Explorer: Science, Observatories, and Community}},  \href{http://arxiv.org/abs/2109.09882}{{\tt arXiv:2109.09882}}.

\bibitem{Colpi:2024xhw}
M.~Colpi et~al., {\it {LISA Definition Study Report}},  \href{http://arxiv.org/abs/2402.07571}{{\tt arXiv:2402.07571}}.

\bibitem{TianQin:2015yph}
{\bf TianQin} Collaboration, J.~Luo et~al., {\it {TianQin: a space-borne gravitational wave detector}},  {\em Class. Quant. Grav.} {\bf 33} (2016), no.~3 035010, [\href{http://arxiv.org/abs/1512.02076}{{\tt arXiv:1512.02076}}].

\bibitem{Hu:2017mde}
W.-R. Hu and Y.-L. Wu, {\it {The Taiji Program in Space for gravitational wave physics and the nature of gravity}},  {\em Natl. Sci. Rev.} {\bf 4} (2017), no.~5 685--686.

\bibitem{penrose1965gravitational}
R.~Penrose, {\it Gravitational collapse and space-time singularities},  {\em Physical Review Letters} {\bf 14} (1965), no.~3 57.

\bibitem{hawking1970singularities}
S.~W. Hawking and R.~Penrose, {\it The singularities of gravitational collapse and cosmology},  {\em Proceedings of the Royal Society of London. A. Mathematical and Physical Sciences} {\bf 314} (1970), no.~1519 529--548.

\bibitem{brady1995black}
P.~R. Brady and J.~D. Smith, {\it Black hole singularities: a numerical approach},  {\em Physical review letters} {\bf 75} (1995), no.~7 1256.

\bibitem{hawking1974black}
S.~W. Hawking, {\it Black hole explosions?},  {\em Nature} {\bf 248} (1974), no.~5443 30--31.

\bibitem{hawking2005black}
S.~Hawking, {\it Black holes and the information paradox},  in {\em General Relativity and Gravitation}, pp.~56--62.
\newblock World Scientific, 2005.

\bibitem{bekenstein2004black}
J.~D. Bekenstein, {\it Black holes and information theory},  {\em Contemporary Physics} {\bf 45} (2004), no.~1 31--43.

\bibitem{landgren2019information}
F.~Landgren, {\it The information paradox, a modern review}, .

\bibitem{thiemann2008modern}
T.~Thiemann, {\em Modern canonical quantum general relativity}.
\newblock Cambridge University Press, 2008.

\bibitem{rovelli2004quantum}
C.~Rovelli, {\em Quantum gravity}.
\newblock Cambridge university press, 2004.

\bibitem{han2007fundamental}
M.~Han, Y.~Ma, and W.~Huang, {\it Fundamental structure of loop quantum gravity},  {\em International Journal of Modern Physics D} {\bf 16} (2007), no.~09 1397--1474.

\bibitem{thiemann2003lectures}
T.~Thiemann, {\it Lectures on loop quantum gravity},  in {\em Quantum gravity: From theory to experimental search}, pp.~41--135.
\newblock Springer, 2003.

\bibitem{ashtekar2004background}
A.~Ashtekar and J.~Lewandowski, {\it Background independent quantum gravity: a status report},  {\em Classical and Quantum Gravity} {\bf 21} (2004), no.~15 R53.

\bibitem{giesel2012classical}
K.~Giesel and H.~Sahlmann, {\it From classical to quantum gravity: introduction to loop quantum gravity},  {\em arXiv preprint arXiv:1203.2733} (2012).

\bibitem{rovelli2011zakopane}
C.~Rovelli, {\it Zakopane lectures on loop gravity},  {\em arXiv preprint arXiv:1102.3660} (2011).

\bibitem{perez2013spin}
A.~Perez, {\it The spin-foam approach to quantum gravity},  {\em Living Reviews in Relativity} {\bf 16} (2013) 1--128.

\bibitem{husain2022quantum}
V.~Husain, J.~G. Kelly, R.~Santacruz, and E.~Wilson-Ewing, {\it Quantum gravity of dust collapse: shock waves from black holes},  {\em Physical Review Letters} {\bf 128} (2022), no.~12 121301.

\bibitem{husain2022fate}
V.~Husain, J.~G. Kelly, R.~Santacruz, and E.~Wilson-Ewing, {\it Fate of quantum black holes},  {\em Physical Review D} {\bf 106} (2022), no.~2 024014.

\bibitem{rovelli1996black}
C.~Rovelli, {\it Black hole entropy from loop quantum gravity},  {\em Physical Review Letters} {\bf 77} (1996), no.~16 3288.

\bibitem{perez2017black}
A.~Perez, {\it Black holes in loop quantum gravity},  {\em Reports on Progress in Physics} {\bf 80} (2017), no.~12 126901.

\bibitem{rovelli2014planck}
C.~Rovelli and F.~Vidotto, {\it Planck stars},  {\em International Journal of Modern Physics D} {\bf 23} (2014), no.~12 1442026.

\bibitem{haggard2015quantum}
H.~M. Haggard and C.~Rovelli, {\it Quantum-gravity effects outside the horizon spark black to white hole tunneling},  {\em Physical Review D} {\bf 92} (2015), no.~10 104020.

\bibitem{de2016improved}
T.~De~Lorenzo and A.~Perez, {\it Improved black hole fireworks: Asymmetric black-hole-to-white-hole tunneling scenario},  {\em Physical Review D} {\bf 93} (2016), no.~12 124018.

\bibitem{christodoulou2016planck}
M.~Christodoulou, C.~Rovelli, S.~Speziale, and I.~Vilensky, {\it Planck star tunneling time: An astrophysically relevant observable from background-free quantum gravity},  {\em Physical Review D} {\bf 94} (2016), no.~8 084035.

\bibitem{bianchi2018white}
E.~Bianchi, M.~Christodoulou, F.~d’Ambrosio, H.~M. Haggard, and C.~Rovelli, {\it White holes as remnants: a surprising scenario for the end of a black hole},  {\em Classical and Quantum Gravity} {\bf 35} (2018), no.~22 225003.

\bibitem{PhysRevD.103.106014}
F.~D'Ambrosio, M.~Christodoulou, P.~Martin-Dussaud, C.~Rovelli, and F.~Soltani, {\it End of a black hole's evaporation},  {\em Phys. Rev. D} {\bf 103} (May, 2021) 106014.

\bibitem{soltani2021end}
F.~Soltani, C.~Rovelli, and P.~Martin-Dussaud, {\it End of a black hole’s evaporation. ii.},  {\em Physical Review D} {\bf 104} (2021), no.~6 066015.

\bibitem{rignon2022black}
A.~Rignon-Bret and C.~Rovelli, {\it Black to white transition of a charged black hole},  {\em Physical Review D} {\bf 105} (2022), no.~8 086003.

\bibitem{han2023geometry}
M.~Han, C.~Rovelli, and F.~Soltani, {\it Geometry of the black-to-white hole transition within a single asymptotic region},  {\em Physical Review D} {\bf 107} (2023), no.~6 064011.

\bibitem{oppenheimer1939continued}
J.~R. Oppenheimer and H.~Snyder, {\it On continued gravitational contraction},  {\em Physical Review} {\bf 56} (1939), no.~5 455.

\bibitem{lewandowski2023quantum}
J.~Lewandowski, Y.~Ma, J.~Yang, and C.~Zhang, {\it Quantum oppenheimer-snyder and swiss cheese models},  {\em Physical Review Letters} {\bf 130} (2023), no.~10 101501.

\bibitem{yang2023shadow}
J.~Yang, C.~Zhang, and Y.~Ma, {\it Shadow and stability of quantum-corrected black holes},  {\em The European Physical Journal C} {\bf 83} (2023), no.~7 619.

\bibitem{stashko2024quasinormal}
O.~Stashko, {\it Quasinormal modes and gray-body factors of regular black holes in asymptotically safe gravity},  {\em Physical Review D} {\bf 110} (2024), no.~8 084016.

\bibitem{zhang2023black}
C.~Zhang, Y.~Ma, and J.~Yang, {\it Black hole image encoding quantum gravity information},  {\em Physical Review D} {\bf 108} (2023), no.~10 104004.

\bibitem{gong2024quasinormal}
H.~Gong, S.~Li, D.~Zhang, G.~Fu, and J.-P. Wu, {\it Quasinormal modes of quantum-corrected black holes},  {\em Physical Review D} {\bf 110} (2024), no.~4 044040.

\bibitem{yang2024gravitational}
S.~Yang, Y.-P. Zhang, T.~Zhu, L.~Zhao, and Y.-X. Liu, {\it Gravitational waveforms from periodic orbits around a quantum-corrected black hole},  {\em arXiv preprint arXiv:2407.00283} (2024).

\bibitem{liu2024gravitational}
H.~Liu, M.-Y. Lai, X.-Y. Pan, H.~Huang, and D.-C. Zou, {\it Gravitational lensing effect of black holes in effective quantum gravity},  {\em Physical Review D} {\bf 110} (2024), no.~10 104039.

\bibitem{zi2024eccentric}
T.~Zi and S.~Kumar, {\it Eccentric extreme mass-ratio inspirals: A gateway to probe quantum gravity effects},  {\em arXiv preprint arXiv:2409.17765} (2024).

\bibitem{kantowski1966some}
R.~Kantowski and R.~K. Sachs, {\it Some spatially homogeneous anisotropic relativistic cosmological models},  {\em Journal of Mathematical Physics} {\bf 7} (1966), no.~3 443--446.

\bibitem{Joe:2015jis}
A.~Joe, {\em {Anisotropic Spacetimes and Black Hole Interiors in Loop Quantum Gravity}}.
\newblock PhD thesis, Louisiana State U., 2015.

\bibitem{modesto2006kantowski}
L.~Modesto, {\it The kantowski-sachs space-time in loop quantum gravity},  {\em International Journal of Theoretical Physics} {\bf 45} (2006), no.~12 2235--2246.

\bibitem{casadio2024cosmology}
R.~Casadio, A.~Kamenshchik, and J.~Ovalle, {\it Cosmology from schwarzschild black hole revisited},  {\em Physical Review D} {\bf 110} (2024), no.~4 044001.

\bibitem{hawking1975particle}
S.~W. Hawking, {\it Particle creation by black holes},  {\em Communications in mathematical physics} {\bf 43} (1975), no.~3 199--220.

\bibitem{gibbons1977action}
G.~W. Gibbons and S.~W. Hawking, {\it Action integrals and partition functions in quantum gravity},  {\em Physical Review D} {\bf 15} (1977), no.~10 2752.

\bibitem{parikh2000hawking}
M.~K. Parikh and F.~Wilczek, {\it Hawking radiation as tunneling},  {\em Physical review letters} {\bf 85} (2000), no.~24 5042.

\bibitem{parikh2004secret}
M.~Parikh, {\it A secret tunnel through the horizon},  {\em International Journal of Modern Physics D} {\bf 13} (2004), no.~10 2351--2354.

\bibitem{rahman2012hawking}
M.~A. Rahman and M.~I. Hossain, {\it Hawking radiation of schwarzschild--de sitter black hole by hamilton--jacobi method},  {\em Physics Letters B} {\bf 712} (2012), no.~1-2 1--5.

\bibitem{kim2016schwinger}
S.~P. Kim, {\it Schwinger effect, hawking radiation and unruh effect},  {\em International Journal of Modern Physics D} {\bf 25} (2016), no.~13 1645005.

\bibitem{hooft1985quantum}
G.~Hooft, {\it On the quantum structure of a black hole},  {\em Nuclear Physics B} {\bf 256} (1985) 727--745.

\bibitem{emparan2006black}
R.~Emparan, {\it Black hole entropy as entanglement entropy: a holographic derivation},  {\em Journal of High Energy Physics} {\bf 2006} (2006), no.~06 012.

\bibitem{almheiri2021entropy}
A.~Almheiri, T.~Hartman, J.~Maldacena, E.~Shaghoulian, and A.~Tajdini, {\it The entropy of hawking radiation},  {\em Reviews of Modern Physics} {\bf 93} (2021), no.~3 035002.

\bibitem{Zhang:2005wn}
J.-Y. Zhang and Z.~Zhao, {\it {Hawking radiation via tunneling from Kerr black holes}},  {\em Mod. Phys. Lett. A} {\bf 20} (2005) 1673--1681.

\bibitem{Zhang:2005sf}
J.-Y. Zhang and Z.~Zhao, {\it {Massive particles-prime black hole tunneling and de Sitter tunneling}},  {\em Nucl. Phys. B} {\bf 725} (2005) 173--180.

\bibitem{ghosh2005log}
A.~Ghosh and P.~Mitra, {\it Log correction to the black hole area law},  {\em Physical Review D—Particles, Fields, Gravitation, and Cosmology} {\bf 71} (2005), no.~2 027502.

\bibitem{kaul2000logarithmic}
R.~K. Kaul and P.~Majumdar, {\it Logarithmic correction to the bekenstein-hawking entropy},  {\em Physical Review Letters} {\bf 84} (2000), no.~23 5255.

\bibitem{domagala2004black}
M.~Domagala and J.~Lewandowski, {\it Black-hole entropy from quantum geometry},  {\em Classical and Quantum Gravity} {\bf 21} (2004), no.~22 5233.

\bibitem{meissner2004black}
K.~A. Meissner, {\it Black-hole entropy in loop quantum gravity},  {\em Classical and Quantum Gravity} {\bf 21} (2004), no.~22 5245.

\bibitem{chatterjee2004universal}
A.~Chatterjee and P.~Majumdar, {\it Universal canonical black hole entropy},  {\em Physical Review Letters} {\bf 92} (2004), no.~14 141301.

\bibitem{medved2004comment}
A.~Medved, {\it A comment on black hole entropy or does nature abhor a logarithm?},  {\em Classical and Quantum Gravity} {\bf 22} (2004), no.~1 133.

\bibitem{lin2024effective}
J.~Lin and X.~Zhang, {\it Effective four-dimensional loop quantum black hole with a cosmological constant},  {\em Physical Review D} {\bf 110} (2024), no.~2 026002.

\bibitem{shi2024higher}
Z.~Shi, X.~Zhang, and Y.~Ma, {\it Higher-dimensional quantum oppenheimer-snyder model},  {\em Physical Review D} {\bf 110} (2024), no.~10 104074.

\bibitem{banerjee2008quantum}
R.~Banerjee and B.~R. Majhi, {\it Quantum tunneling and back reaction},  {\em Physics Letters B} {\bf 662} (2008), no.~1 62--65.

\bibitem{banerjee2008quantum2}
R.~Banerjee and B.~R. Majhi, {\it Quantum tunneling beyond semiclassical approximation},  {\em Journal of High Energy Physics} {\bf 2008} (2008), no.~06 095.

\bibitem{banerjee2009quantum}
R.~Banerjee and B.~R. Majhi, {\it Quantum tunneling and trace anomaly},  {\em Physics Letters B} {\bf 674} (2009), no.~3 218--222.

\bibitem{majhi2009fermion}
B.~R. Majhi, {\it Fermion tunneling beyond semiclassical approximation},  {\em Physical Review D—Particles, Fields, Gravitation, and Cosmology} {\bf 79} (2009), no.~4 044005.

\bibitem{majhi2010hawking}
B.~R. Majhi and S.~Samanta, {\it Hawking radiation due to photon and gravitino tunneling},  {\em Annals of Physics} {\bf 325} (2010), no.~11 2410--2424.

\bibitem{ashtekar2006quantum}
A.~Ashtekar, T.~Pawlowski, and P.~Singh, {\it Quantum nature of the big bang},  {\em Physical review letters} {\bf 96} (2006), no.~14 141301.

\bibitem{yang2009alternative}
J.~Yang, Y.~Ding, and Y.~Ma, {\it Alternative quantization of the hamiltonian in loop quantum cosmology},  {\em Physics Letters B} {\bf 682} (2009), no.~1 1--7.

\bibitem{assanioussi2018t}
M.~Assanioussi, A.~Dapor, and K.~Liegener, {\it T. paw lowski, emergent de sitter epoch of the quantum cosmos},  {\em Phys. Rev. Lett} {\bf 121} (2018) 081303.

\bibitem{israel1966singular}
W.~Israel, {\it Singular hypersurfaces and thin shells in general relativity},  {\em Il Nuovo Cimento B (1965-1970)} {\bf 44} (1966), no.~1 1--14.

\bibitem{painleve1921mecanique}
P.~Painlev{\'e}, {\it La m{\'e}canique classique et la th{\'e}orie de la relativit{\'e}},  {\em Comptes Rendus Academie des Sciences (serie non specifiee)} {\bf 173} (1921) 677--680.

\bibitem{cafaro2024status}
L.~Cafaro and J.~Lewandowski, {\it Status of birkhoff’s theorem in the polymerized semiclassical regime of loop quantum gravity},  {\em Physical Review D} {\bf 110} (2024), no.~2 024072.

\bibitem{singha2022thermodynamics}
C.~Singha, {\it Thermodynamics of multi-horizon spacetimes},  {\em General Relativity and Gravitation} {\bf 54} (2022), no.~4 38.

\bibitem{saghafi2023hawking}
S.~Saghafi and K.~Nozari, {\it Hawking-like radiation as tunneling from a cosmological black hole in modified gravity: semiclassical approximation and beyond},  {\em General Relativity and Gravitation} {\bf 55} (2023), no.~1 20.

\bibitem{nozari2007existence}
K.~Nozari and A.~Sefidgar, {\it On the existence of the logarithmic correction term in black hole entropy-area relation},  {\em General Relativity and Gravitation} {\bf 39} (2007) 501--509.

\bibitem{snyder1947quantized}
H.~S. Snyder, {\it Quantized space-time},  {\em Physical Review} {\bf 71} (1947), no.~1 38.

\bibitem{seiberg1999string}
N.~Seiberg and E.~Witten, {\it String theory and noncommutative geometry},  {\em Journal of High Energy Physics} {\bf 1999} (1999), no.~09 032.

\bibitem{douglas2001noncommutative}
M.~R. Douglas and N.~A. Nekrasov, {\it Noncommutative field theory},  {\em Reviews of Modern Physics} {\bf 73} (2001), no.~4 977.

\bibitem{aschieri2005gravity}
P.~Aschieri, C.~Blohmann, M.~Dimitrijevi{\'c}, F.~Meyer, P.~Schupp, and J.~Wess, {\it A gravity theory on noncommutative spaces},  {\em Classical and Quantum Gravity} {\bf 22} (2005), no.~17 3511.

\bibitem{nozari2008hawking}
K.~Nozari and S.~H. Mehdipour, {\it Hawking radiation as quantum tunneling from a noncommutative schwarzschild black hole},  {\em Classical and Quantum Gravity} {\bf 25} (2008), no.~17 175015.

\bibitem{amelino2006black}
G.~Amelino-Camelia, M.~Arzano, Y.~Ling, and G.~Mandanici, {\it Black-hole thermodynamics with modified dispersion relations and generalized uncertainty principles},  {\em Classical and Quantum Gravity} {\bf 23} (2006), no.~7 2585.

\bibitem{nozari2006comparison}
K.~Nozari and A.~Sefidgar, {\it Comparison of approaches to quantum correction of black hole thermodynamics},  {\em Physics Letters B} {\bf 635} (2006), no.~2-3 156--160.

\bibitem{nozari2012natural}
K.~Nozari and S.~Saghafi, {\it Natural cutoffs and quantum tunneling from black hole horizon},  {\em Journal of High Energy Physics} {\bf 2012} (2012), no.~11 1--18.

\bibitem{ashtekar2003quantum}
A.~Ashtekar, S.~Fairhurst, and J.~L. Willis, {\it Quantum gravity, shadow states and quantum mechanics},  {\em Classical and Quantum Gravity} {\bf 20} (2003), no.~6 1031.

\bibitem{corichi2007polymer}
A.~Corichi, T.~Vuka{\v{s}}inac, and J.~A. Zapata, {\it Polymer quantum mechanics and its continuum limit},  {\em Physical Review D—Particles, Fields, Gravitation, and Cosmology} {\bf 76} (2007), no.~4 044016.

\bibitem{gorji2014polymeric}
M.~Gorji, K.~Nozari, and B.~Vakili, {\it Polymeric quantization and black hole thermodynamics},  {\em Physics Letters B} {\bf 735} (2014) 62--68.

\bibitem{penington2020entanglement}
G.~Penington, {\it Entanglement wedge reconstruction and the information paradox},  {\em Journal of High Energy Physics} {\bf 2020} (2020), no.~9 1--84.

\bibitem{almheiri2019entropy}
A.~Almheiri, N.~Engelhardt, D.~Marolf, and H.~Maxfield, {\it The entropy of bulk quantum fields and the entanglement wedge of an evaporating black hole},  {\em Journal of High Energy Physics} {\bf 2019} (2019), no.~12 1--47.

\bibitem{almheiri2020page}
A.~Almheiri, R.~Mahajan, J.~Maldacena, and Y.~Zhao, {\it The page curve of hawking radiation from semiclassical geometry},  {\em Journal of High Energy Physics} {\bf 2020} (2020), no.~3 1--24.

\bibitem{almheiri2019islands}
A.~Almheiri, R.~Mahajan, and J.~Maldacena, {\it Islands outside the horizon},  {\em arXiv preprint arXiv:1910.11077} (2019).

\bibitem{hashimoto2020islands}
K.~Hashimoto, N.~Iizuka, and Y.~Matsuo, {\it Islands in schwarzschild black holes},  {\em Journal of High Energy Physics} {\bf 2020} (2020), no.~6 1--21.

\bibitem{ryu2006aspects}
S.~Ryu and T.~Takayanagi, {\it Aspects of holographic entanglement entropy},  {\em Journal of High Energy Physics} {\bf 2006} (2006), no.~08 045.

\bibitem{hubeny2007covariant}
V.~E. Hubeny, M.~Rangamani, and T.~Takayanagi, {\it A covariant holographic entanglement entropy proposal},  {\em Journal of High Energy Physics} {\bf 2007} (2007), no.~07 062.

\bibitem{engelhardt2015quantum}
N.~Engelhardt and A.~C. Wall, {\it Quantum extremal surfaces: holographic entanglement entropy beyond the classical regime},  {\em Journal of High Energy Physics} {\bf 2015} (2015), no.~1 1--27.

\bibitem{page1993information}
D.~N. Page, {\it Information in black hole radiation},  {\em Physical review letters} {\bf 71} (1993), no.~23 3743.

\bibitem{page2013time}
D.~N. Page, {\it Time dependence of hawking radiation entropy},  {\em Journal of Cosmology and Astroparticle Physics} {\bf 2013} (2013), no.~09 028.

\end{thebibliography}\endgroup
\end{document}